\RequirePackage{fix-cm}
\documentclass[smallextended]{svjour3}       
\smartqed  
\usepackage{graphicx}
\begin{document}

\title{Hall resistivity correlations in disordered electron-doped Nd$_{2-x}$Ce$_x$CuO$_{4+\delta}$ films
}


\author{T.Charikova         \and
        N.Shelushinina  \and
				G.Harus  \and
				D.Petukhov \and
				O.Petukhova \and
				A.Ivanov}


\institute{T.Charikova \at
             1. M.N.Mikheev Institute of Metal Physics Ural Branch of RAS (IMP UB RAS), Ekaterinburg, Russia,  2. Ural Federal University, Ekaterinburg, Russia\\
              Tel.: +7-343-3783733\\
             \email{charikova@imp.uran.ru}           
           \and
           N.Shelushinina \at
             IMP UB RAS, Ekaterinburg, Russia,\email{shel@imp.uran.ru}      
						 \and
						G.Harus \at
						IMP UB RAS,  \email{kharus@imp.uran.ru}       
						\and
						D.Petukhov \at
					IMP UB RAS,
						              \email{dpetukhov@imp.uran.ru}       
						\and
						O.Petukhova \at
						IMP UB RAS,
						              \email{oep@yandex.ru}       
						\and
						A.Ivanov \at
						National Research Nuclear University MEPhI, Moscow, Russia,
						              \email{andrej.ivanov@gmail.com}       
}

\date{Received: date / Accepted: date}

\maketitle

\begin{abstract}
The resistivity tensor correlations $\rho_{xy}$($B$)$\sim$ $[\rho_{xx}(B)]^\beta$ for the mixed state magnetic field dependencies of the resistivity tensor of electron-doped Nd$_{2-x}$Ce$_x$CuO$_{4+\delta}$/SrTiO$_3$ single crystal films near the antiferromagnetic (AF) - superconducting (SC) phase transition and with varying degree of disorder ($\delta$) were studied. The decrease of $\beta$ from 1.2$\pm$0.2 at $x$ = 0.14 to 0.6$\pm$0.1 at $x$ = 0.15 points out on the evidence of the change from the anisotropic $s$ - wave to the $d$ - wave pairing symmetry in the external magnetic field at the transition from underdoped to optimally doped region. Peculiarities of the power law dependence of the vortex motion in the mixed state can be connected with some features of the nonstoichiometric disorder in layered electron-doped superconductors.
\end{abstract}

\section{Introduction}
\label{intro}

The attempts to understand the reasons of the appearance of the superconductivity (SC) lead to the investigations of the quantum phase transition either from the temperature or from the magnetic field. The compounds both with electron- and with hole-doping represent permanent interest. Asymmetry between physical properties of the  electron- and hole-doped cuprate SC's  is owing to the peculiarities of the crystal structures of a parent compounds \cite{1}. Insulating behaviour of the parent compound is due to a long-range magnetic order for electron-doped materials. In the case of the hole-doped materials the insulating behaviour depends on the strong electronic correlations. Electron doping is generated by replacing the Nd$^{3+}$ ions in the parent compound Nd$_2$CuO$_4$ with Ce$^{4+}$ to form a non-superconducting antiferromagnet Nd$_{2-x}$Ce$_x$CuO$_{4+\delta}$ \cite{2}. Additional annealing in oxygen-free atmosphere leads to suppress the static AF order and  to appearance of the superconductivity \cite{3,4,5,6}. During the annealing process a certain amount of the oxygen excess is removed and the lattice structure isn't practically destroyed  \cite{4,7,8}.  This gives us a chance to get the compounds with different nonstoichiometric disorder ($\delta$) and to investigate the features of magnetotransport properties near the transition from AF to SC. Some studies of electron-doped systems phase diagram indicate that the pseudogap state is hidden under the AF and SC regions \cite{9} unlike of the hole-doped compound phase diagram. An important topic of discussion of the features in electron-doped compounds is a competition of $d$-wave SC and AF order and the presence or absence of spin- or charge- density wave phases in underdoped region\cite{10,11}. Exactly in this range in n-doped cuprates a Fermi surface change from electron-like pockets to a  coexisting of a small hole-like  and  electron-like pockets is observed\cite{3,12,13}. All of these facts motivate us to investigate the dependencies of the electrical $\rho_{xx}$($B$) and Hall $\rho_{xy}$($B$) resistivity in external magnetic field in electron-doped Nd$_{2-x}$Ce$_x$CuO$_{4+\delta}$ with nonstochiometric disorder ($\delta$) in order to analyze the conditions for the appearance of the Hall and dissipative resistivity correlations in the presence of evolution from AF - (underdoped region) to SC - order (optimally doped region).

\section{Experimental details, results and discussion} 

\label{sec:1}

The series of Nd$_{2-x}$Ce$_x$CuO$_{4+\delta}$/SrTiO$_3$ epitaxial films ($x$ = 0.14,  0.15) with the standard (001) orientation were synthesized by the pulsed laser deposition \cite{5}. Then the films were subjected to heat treatment (annealing) under various conditions to obtain samples with various oxygen content.  As a result, three types of samples with $x$=0.15 were obtained: ``as grown'' samples, ``optimally reduced'' - after annealing in a vacuum and ``non optimally reduced'' samples \cite{14}.  For films with $x$ = 0.14 were obtained two types of samples: ``optimally reduced''  and two samples ``non optimally reduced'' - non optimally annealed in vacuum \cite{14,15}. The thickness of the films was $d$ = 1600-3800 {\AA}.

Our previous investigations point out that the change in the oxygen concentration leads to a change in the degree of nonstoichiometric disorder \cite{6,7,8,15}. Annealing in an oxygen-free atmosphere leads to a change in impurity scattering, while having little effect on the concentration of charge carriers and we will use the concepts of the theory of random two-dimensional systems where the parameter $k_F$$\ell$ serves as the measure of disorder in the system. 

We have made the resistivity tensor $\rho_{xx}$($B$) and $\rho_{xy}$($B$) investigations as a function of the external magnetic field (I$\Vert$ab, B$\Vert$c) in the temperature range $T$ = (0.4 - 40)K  in a mixed and normal states of electron - doped high-temperature superconductor Nd$_{2-x}$Ce$_x$CuO$_{4+\delta}$  ($x$ = 0.14 and 0.15) with varying degree of nonstoichiometric disorder $\delta$. Temperature dependencies of the resistivity were measured at the magnetic field up to $B$ = 9 T using the Quantum Design PPMS 9 in the temperature range $T =$ (1.8 $\div$ 300) K. Magnetic field dependencies of the resistivity measurements in the temperature $T$ = (0.4 - 4.2) K were performed in the solenoid "Oxford Instruments" in magnetic fields up to $B$ = 12 T (Institute of Metal Physics RAS, Ekaterinburg). Temperature dependence of the a.c. magnetic susceptibility was measured using Quantum Design SQUID magnetometer MPMS 5 at the frequence of 81 Hz and with a.c. magnetic field amplitude of 4$\cdot$10$^{-4}$ T.

The three types of the optimally doped ($x$ = 0.15) SC films had a following parameters:

1. ``As grown'' film  - hasn't SC transition at $B$ = 0,  $k_F$$\ell$ = 8.6;

2. ``Non optimally reduced'' film - $T_c$ = 14.9 K,  $k_F$$\ell$ = 9.1;

3. ``Optimally reduced'' film - $T_c$ = 22 K,  $k_F$$\ell$ = 51.6.

Increase of the nonstoichiometric disorder leads to the rapidly decrease of the critical temperature. The observed dependence $T_c$ from $k_F$$\ell$ (which extracted as $k_F\ell$ = $hc_0$/$e^2 \rho_{xx}$) is associated in high-temperature superconductors with the manifestation of the d-wave pairing symmetry \cite{14}. 

Temperature dependencies of the resistivity in external magnetic field up to  $B$ = 9 T for these types of the films were presented in \cite{14}. In optimally - and non optimally reduced films we have found the sharp SC transition in zero magnetic field. As the magnetic field increases the SC transition moves to lower temperatures without considerable change of the transition width. The picture destroys only for as grown samples: the transition broadens and disappears. 

\begin{figure}
\begin{minipage}{0.49\linewidth}
\center{\includegraphics{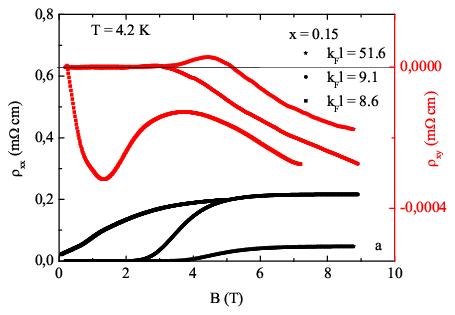}}
\end{minipage}
\hfill
\begin{minipage}{0.49\linewidth}
\center{\includegraphics{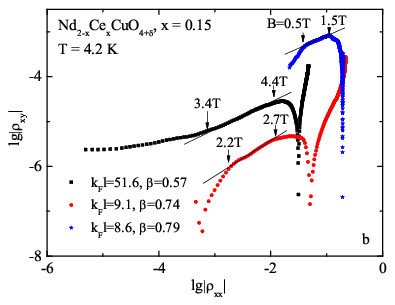}}
\end{minipage}
\caption{a). Magnetic field dependencies of the longitudinal resistivity $\rho_{xx}$$(B)$  and Hall resistivity $\rho_{xy}$$(B)$ for single crystal films of optimally doped ($x$ = 0.15) Nd$_{2-x}$Ce$_x$CuO$_{4+\delta}$ with different disorder parameters at $T$ = 4.2 K: (Color figure online.)
b). Log-log plot of the Hall resistivity $\rho_{xy}$ vs longitudinal resistivity $\rho_{xx}$  of optimally doped ($x$ = 0.15) Nd$_{2-x}$Ce$_x$CuO$_{4+\delta}$/SrTiO$_3$ films with different disorder parameters at $T$ = 4.2 K: (Color figure online.)}
\label{Fig1}
\end{figure}

The magnetic field dependencies of the longitudinal and Hall resistivities of optimally doped ($x$ = 0.15) superconductor Nd$_{2-x}$Ce$_x$CuO$_{4+\delta}$ with different disorder parameters $k_F$$\ell$ at $T$ = 4.2 K are shown in Fig.~\ref{Fig1}a. In the mixed state the longitudinal and Hall resistivities there are the resistivity correlations $\rho_{xy}$($B$)$\sim$ $[\rho_{xx}(B)]^\beta$ that was found for different kinds of HTSC (see references in\cite{15}). The log-log plot for Nd$_{1.85}$Ce$_{0.15}$CuO$_{4+\delta}$/SrTiO$_3$ films with different disorder parameters $k_F$$\ell$ at $T$ = 4.2K (Fig.~\ref{Fig1}b) is presented. For ``optimally reduced'' film index $\beta$ is equal 0.57 $\pm$ 0.03 and increases with the increase of the disorder up to $\beta$ = 0.74 $\pm$ 0.03 for the film with $k_F$$\ell$ = 9.1 and $\beta$ = 0.79 $\pm$ 0.03 for the film with $k_F$$\ell$ = 8.6. The similar values of the index $\beta$ = 0.8$\pm$ 0.2 for optimal doped single crystal Nd$_{2-x}$Ce$_x$CuO$_{4}$ were presented in \cite{16}. Acoording to \cite{17} the correlation for $\rho_{xy}$($B$) and $\rho_{xx}(B)$ with $\beta$$\sim$1 may appear in the mixed state of $d$-wave SC.  Figure ~\ref{Fig1}b shows the range of the magnetic field, where there are a stable linear dependence of the power law. This behavior does not depend on the sign of the Hall resistivity. It should be noted that Hall resistivity correlations appear at the magnetic field where we begin to detect the resistivity. Therefore these correlations may be associated with features of the flux flow regime with the increasing of the magnetic field. 

In underdoped region ($x$ = 0.14) we have two types of SC films with a following parameters:

1. ``Non optimally reduced'' films - $T_c$ = 17.6 K,  $k_F$$\ell$ = 6.0 and $T_c$ = 18.1 K,  $k_F$$\ell$ = 2.3;

2. ``Optimally reduced'' film - $T_c$ = 18.8 K,  $k_F$$\ell$ = 10.8.

In these underdoped films the critical temperature remains unchanged with the increase of the nonstochiometric disorder that may indicate a some change of the pairing symmetry (for example, assymetric $s$-wave pairing \cite{14}). 

On the assumption of the possible coexistence of AF and SC we have obtained the a.c.susceptibility data (both $\chi'$ and $\chi''$) for underdoped non optimally reduced single crystal film Nd$_{1.86}$Ce$_{0.14}$CuO$_{4+\delta}$/SrTiO$_3$ with $k_F \ell$ = 6.0 (\cite{14}). The onset of diamagnetic response occurs at $T_c \simeq$ 18.8 K. At the same temperature the sufficiently sharp resistivity SC transition is observed. But there is not fairly fast drop in $\chi'$. This feature can be also seen in the $\chi''$($T$) behavior: the a.c. losses increases over the all temperature range below $T_c^{onset}$. This fact indicates the possible existence of the AF regions, where the diamagnetic response doesn't exist. However, along with this the transport percolation leads to the SC transition.

\begin{figure}
  \includegraphics{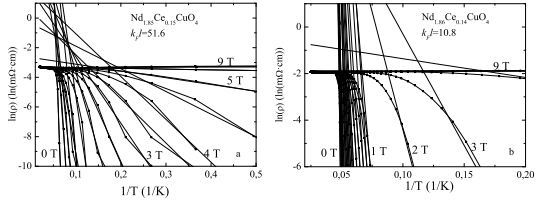}
\caption{Arrenius plots of the longitudinal resistivity $\rho(T)$ at different external magnetic field for optimally doped (a) and underdoped (b) optimally reduced films.}
\label{Fig2}       
\end{figure}

The magnetic field dependencies of  $\rho_{xx}$$(B)$ and $\rho_{xy}$$(B)$ of underdoped ($x$ = 0.14) superconductor Nd$_{2-x}$Ce$_x$CuO$_{4+\delta}$ with varying $k_F$$\ell$ at $T$ = 4.2 K were investigated and presented in \cite{15}. As in the case of the optimally doped compounds we see the increase of the index $\beta$ with increase of the disorder ($k_F$$\ell$ decreases) from $\beta$ = 1.11 $\pm$ 0.01 up to $\beta$ = 1.21 $\pm$ 0.01 for the film with $k_F$$\ell$ = 6.0 and $\beta$ = 1.17 $\pm$ 0.03 for the film with $k_F$$\ell$ = 2.3. So, in underdoped region the index $\beta$ is twice as in optimally doped region. The existence of AF order in this region may affect the symmetry of the SC order parameter and appear to change the value of the index $\beta$. It should be emphasized that the disorder parameter of the optimally reduced underdoped film is practically equal to the value of $k_F$$\ell$ for the ``as grown'' optimally doped film: the disorder decreases at the initial stage with the increasing of doping to the optimum, further annealing process reduces the degree of disorder. This makes it possible to analyze the features of the beginning of the transition to the normal state in the presence of an electric current and magnetic field (Fig. ~\ref{Fig2}). 

\begin{figure}
  \includegraphics{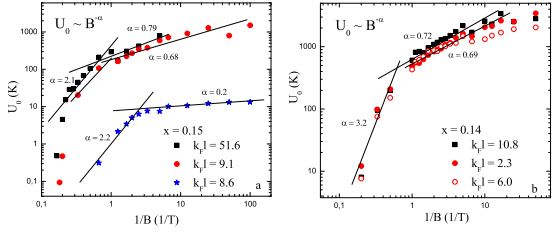}
\caption{Activation energy $U_0$ as a function of the inverse of the magnetic field for optimally doped (a) and underdoped (b) films with a different degree of disorder: (Color figure online.)}
\label{Fig3}       
\end{figure}

A linear behavior ln$\rho$(1/$T$) is observed for all the investigation films, therefore $\rho(T,B)$ = $\rho_0$exp(-$U(T,B)$/k$_B$$T$) with activation energy $U(T,B)$ = $U_0$($B$)g($t$), where $U_0$($B$) $\sim$ $B^{-\alpha}$ with exponent $\alpha$ that varies by changing of the magnetic field and g($t$) = (1 - $t^2$)(1 - $t^4$)$^{1/2}$ with $t$ = $T/T_c$ \cite{18,19,20}. The activation energy $U_0$ as a function of the inverse of the external magnetic field for the investigated electron-doped films with different degree of disorder is presented in Figure ~\ref{Fig3}. In layered superconductors \cite{21} one can define certain regions of the vortex dynamic with the change of the magnetic field: in low fields - a single vortex pancake pinning, with the increasing of the magnetic field - the regime of the a 2D bundle of pancake vortices becomes pinned collectively. The further increase leads to the Josephson coupling between 2D bundles in neighboring layers. The coupling into the third dimension in high magnetic field becomes dominant, and the collectively pinned object takes the shape of a 3D vortex bundle. 

So, for optimally doped ``as grown'' films activation energy $U_0$ is low and up to $B$ = 0.5T has practically constant behaviour $U_0(B)$ due to a single vortex pinning in the weak-pinning-regime according the nonstoichiometric disorder. In higher magnetic field $U_0$ decreases and resistivity tensor correlations appear (Fig. ~\ref{Fig1}b) that may be related to 3D collectively vortex regime. 

For ``non optimally reduced'' and ``optimally reduced'' films with $x$ = 0.15 the value of $U_0$ in low magnetic field is more than  two orders of magnitude greater than in ``as grown'' film and is observed the behaviour of $U_0(B)$ with $\alpha$ $\sim$ 0.7 (very similar to \cite{20}). This behavior may be connected with 2D collective pinning regime in layered Nd$_{1.85}$Ce$_{0.15}$CuO$_{4+\delta}$ with $\delta$ $\to$ 0 by removing of the apical oxygen. In the magnetic fields $B$ $>$ 2 T (``non optimally reduced'' films) and $B$ $>$ 3 T (``optimally reduced'' films) are observed the changes to the field dependencies of $U_0$ with $\alpha$ $\sim$ 2.1 - 2.2 that may correspond to the 3D collectively vortex regime with the correlations of the longitudinal and Hall resistivities (Fig. ~\ref{Fig1}b). 

The similar picture is observed for underdoped films with the difference degree of disorder. It is interesting that there is no difference in the behavior of $U_0(B)$ in the case of ``non optimally reduced'' and ``optimally reduced'' films. The increase of the nonstoichiometric disorder in underdoped region does not change the characteristics of the motion of the vortex system: at the $B$ $<$ 1T 
occurs 2D collective pinning regime and in the region $B$ $>$ 2 - 3 T appear resistivity tensor correlations and 3D-pinning regime.

\section{Conclusions}

In summary, the resistivity tensor correlations $\rho_{xy}$($B$)$\sim$ $[\rho_{xx}(B)]^\beta$ for the mixed state magnetic field dependencies of the resistivity tensor of electron-doped Nd$_{2-x}$Ce$_x$CuO$_{4+\delta}$/SrTiO$_3$ single crystal films in underdoped and optimally doped regions with varying degree of disorder ($\delta$) were investigated. It was found that the index $\beta$ is equal 0.57 $\pm$ 0.03 for optimally reduced film and increases with the increase of the disorder and twice smaller comparing with underdoped compounds. The existence of the modification of the flux flow regime  with the increase of the magnetic field can lead to a change in the correlation of the longitudinal and Hall resistance at the transition from underdoped to optimally doped regions of Nd$_{2-x}$Ce$_x$CuO$_{4+\delta}$ system that may by connected with the change in the symmetry of the order parameter. 

\begin{acknowledgements}
The research was carried within the state assignment of FASO of Russia (theme "Electron" N. 01201463326), supported in part by the Program of fundamental research of the UB of RAS (project N. 15-8-2-6) with partial support of RFBR (grant N. 15-02-02270).
\end{acknowledgements}

\end{document}